\documentclass[conference]{IEEEtran}
\makeatletter
\def\ps@headings{%
\def\@oddhead{\mbox{}\scriptsize\rightmark \hfil \thepage}%
\def\@evenhead{\scriptsize\thepage \hfil \leftmark\mbox{}}%
\def\@oddfoot{}%
\def\@evenfoot{}}
\makeatother
\usepackage[noend]{algorithmic}
\usepackage{algorithm}
\usepackage[cmex10]{amsmath}
\usepackage{amsthm, amssymb, cite, graphicx, epsfig, url, color, setspace}
\numberwithin{algorithm}{section}
\newcommand{\ignore}[1]{}

\newtheorem{proposition}{Proposition}[section]

\newtheorem{lemma}{Lemma}[section]

\usepackage[tight,footnotesize]{subfigure}

\begin{document}

\title{Maximizing Social Welfare in Operator-based Cognitive Radio Networks under Spectrum Uncertainty and Sensing Inaccuracy}

\author{\IEEEauthorblockN{Shuang Li\IEEEauthorrefmark{1},
Zizhan Zheng\IEEEauthorrefmark{2},
Eylem Ekici\IEEEauthorrefmark{2} and
Ness Shroff\IEEEauthorrefmark{1}\IEEEauthorrefmark{2}}
\IEEEauthorblockA{\IEEEauthorrefmark{1}Department of Computer Science and Engineering\\
 Email: li.908@osu.edu}
\IEEEauthorblockA{\IEEEauthorrefmark{2}Department of Electrical and Computer Engineering\\
The Ohio State University,
Columbus, Ohio 43210\\
Email: \{zhengz, ekici, shroff\}@ece.osu.edu}
}

\maketitle

\begin{abstract}
In Cognitive Radio Networks (CRNs), secondary users (SUs) are allowed to opportunistically access the unused/under-utilized channels of primary users (PUs). To utilize spectrum resources efficiently, an auction scheme is often applied where an operator serves as an auctioneer and accepts spectrum requests from SUs. Most existing works on spectrum auctions assume that the operator has perfect knowledge of PU activities. In practice, however, it is more likely that the operator only has statistical information of the PU traffic when it is trading a spectrum hole, and it is acquiring more accurate information in real time. In this paper, we distinguish PU channels that are under the control of the operator, where accurate channel states are revealed in real-time, and channels that the operator acquires from PUs out of its control, where a sense-before-use paradigm has to be followed. Considering both spectrum uncertainty and sensing inaccuracy, we study the social welfare maximization problem for serving SUs with various levels of delay tolerance. We first model the problem as a finite horizon Markov decision process when the operator knows all spectrum requests in advance, and propose an optimal dynamic programming based algorithm. We then investigate the case when spectrum requests are submitted online, and propose a greedy algorithm that is 1/2-competitive for homogeneous channels and is comparable to the offline algorithm for more general settings. We further show that the online algorithm together with a payment scheme achieves incentive compatibility for the SUs while guaranteeing a non-negative revenue for the operator.
\end{abstract} 
\section{Introduction}

With the ever-growing demand for wireless spectrum, Cognitive Radio Networks (CRNs) have been proposed to better utilize spectrum holes in wireless networks. In CRNs, secondary users (SUs) are allowed to opportunistically access the channels of primary users (PUs).
To utilize spectrum resources efficiently, an auction framework is often applied where an operator serves as an auctioneer and accepts requests from SUs. These frameworks are implemented via a resource allocation and a payment scheme with the objective of maximizing either social welfare or revenue~\cite{Zhou,Wang,Gop,Gao,Chen}.

Most existing works on spectrum auctions, however, assume that the operator has perfect knowledge of PU activities in a given period of time. They ignore the uncertainty of channel states caused by the uncertain and frequent PU usage. Hence, these existing auction schemes are mainly applicable to spectrum resources that tend to be available for relatively long periods of time. For instance, the interval between two adjacent auctions is assumed to be 30 minutes or longer in~\cite{Wang}. However, to allow more efficient spectrum utilization and relieve spectrum congestion, spectrum holes at smaller time scales need to be explored. A straightforward extension of current approaches to this more dynamic environment would require auctions to be conducted frequently, which would incur high communication and management overhead. A more reasonable approach is to again consider a relatively long period of time, where the operator only has statistical information of the PU traffic when trading spectrum holes. More accurate information is acquired later in real-time. \emph{Therefore, an auction scheme that takes spectrum uncertainty into account is needed.} \ignore{In such a dynamic environment, Due to this limitation, spectrum holes at small scale cannot be efficiently utilized. the period of auction only lasts for a time slot or other units of a short period of time so that accurate channel state information can be collected. Since the collection has to be done frequently, high overhead will be incurred.}

To further improve spectrum utilization, besides trading spectrum holes that are fully under the control of the operator, as commonly assumed in the spectrum auction literature, the operator may choose to acquire licensed channels out of its control to further improve social welfare or revenue. To avoid interference with PUs, a {\it sense-before-use} paradigm must be followed in this case. The operator must first identify spectrum holes in a channel, e.g., by coordinating SUs to sense the channel, before allocating the holes to SUs. While spectrum sensing has been extensively studied in the CRN literature~\cite{li-single-channel,li-multi-channel,sensing-survey}, the joint problem of sensing and spectrum auction remains unexplored.

In this paper, we propose a spectrum allocation framework that takes both {\it spectrum uncertainty} and {\it sensing inaccuracy} into account. In particular, we consider two types spectrum resources: PU channels that are under the control of the operator, and the channels that the operator acquires from PUs out of its control. In practice, wireless service providers (WSP) act as operators, and they may cover areas that almost completely overlap. SUs registered with one of them may access spectrum from other WSPs as will be introduced in our model. In both types of channels, PU traffic on each channel is assumed to follow a known i.i.d. Bernoulli distribution. For the first type of channels, the real-time channel state can be learned accurately by the operator. For the second type of channels, a sense-before-use paradigm must be followed, where a collision with the PU traffic due to sensing inaccuracy incurs a penalty.

Using a fixed set of channels of each type, we study the joint spectrum sensing and allocation problem to serve spectrum requests with arbitrary valuations and arbitrary levels of delay tolerance. The objective of the operator is to maximize social welfare, which equals to the valuations obtained from successfully served requests minus the cost due to collisions. We consider both the scenario where the operator knows all spectrum requests in advance, and the setting when spectrum requests are submitted online. While our online setting is similar to the online spectrum auction schemes considered in~\cite{Deek,Xu}, the key difference is that sensing inaccuracy is not considered in these existing works. Hence, the approaches in~\cite{Deek,Xu} can only be applied to cases where accurate real time channel states are obtainable, which is not always the case.

Our contributions can be summarized as follows:

\begin{itemize}
\item We model the joint sensing and spectrum allocation problem as a finite horizon Markov decision process when all spectrum requests are revealed to the operator offline, i.e., ahead of time. We develop an optimal dynamic programming based algorithm, which serves as a baseline for the achievable social welfare.

\item We propose a greedy algorithm for the case when spectrum requests are submitted online. We prove that the online algorithm is 1/2-competitive for homogeneous channels, and we show that it achieves performance comparable to the offline algorithm for more general settings by numerical results.

\item We further extend the online algorithm by introducing a payment scheme to ensure incentive compatibility for SUs while guaranteeing a non-negative revenue for the operator.
\end{itemize}

The paper is organized as follows: The system model and problem formulation are introduced in Section~\ref{sec:model}. Our solutions to the problem with offline and online requests are presented in Sections~\ref{sec:off_wo} and~\ref{sec:on_wo}, respectively. The online auction scheme for ensuring incentive compatibility for SUs and non-negative revenue for the operator is then discussed in Section~\ref{sec:inc}. In Section~\ref{sec:simu}, numerical results are presented to illustrate the performance of the greedy online algorithm in general cases, and the tradeoff between social welfare and revenue. We conclude the paper in Section~\ref{sec:con}. 
\section{System Model and Problem Formulation}\label{sec:model}

\begin{figure}[tb]
    \begin{center}
    \setlength{\unitlength}{1in}
    \includegraphics[scale=0.4]{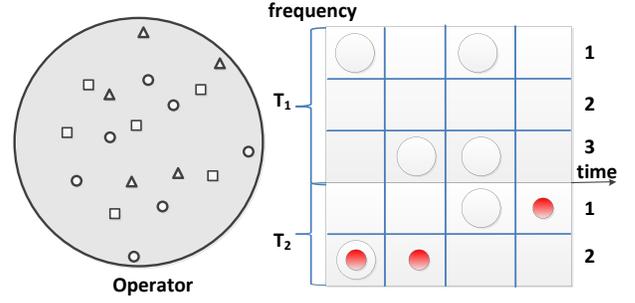}
    \end{center}
\vspace{-1.2em}
\caption{System model of the CRN. In the left figure, small circles are SUs, squares represent PUs registered at the operator, and triangles are PUs out of the operator's control. The big circle is the coverage area of the operator. The right figure shows the availability of channels in $T_1$ and $T_2$. The big circle means the channel is busy and the small red circle means the channel is sensed idle.}
\vspace{-1em}
\label{fig:model}
\end{figure}

We consider a cognitive radio network with a single operator and multiple SUs registered with it (see Figure~\ref{fig:model}). The operator manages multiple orthogonal channels and controls the corresponding network composed of PUs. We focus on downlink transmission at the operator with power control. A time slotted system is considered with all PU and SU transmissions synchronized. All SUs are assumed to be in the interference range of each other and that of PUs, and hence each channel can be assigned to at most one SU at any time when it is not used by PUs.

The spectrum pool consists of two types of channels, those managed by the operator and those that are not. The operator is aware of the downlink activity of its own PUs at the beginning of each time slot. The set of the spectrum bands\footnote{We use channel and spectrum band interchangeably.} managed by the operator is denoted by $T_1$. However, the activities of PUs not managed by the operator are unknown. Bands accessed by these PUs are denoted by $T_2$.
To access bands in $T_2$, SUs cooperatively sense them and report their sensing results to the operator. The operator then makes a fusion decision on the activities of bands in $T_2$ and selects a subset of channels sensed idle to serve the SUs. We only consider the set of PUs located in the coverage area of the operator so that all SUs in the system have the cognitive capability and can sense spectrum in $T_2$. We assume that the sensing cost is low and even negligible. In practice, wireless service providers (WSP) act as operators, and they may cover areas that almost overlap. SUs registered with one of them may access spectrum from other WSPs as introduced in our model.

We assume that the spectrum bands in $T_1$ and $T_2$ have the same capacity, which is normalized to $1$. PU activities on these channels follow an i.i.d. Bernoulli distribution in each time slot. For instance, in Figure~\ref{fig:model}, there are three channels in $T_1$ and two channels in $T_2$. In time slot $1$, channels 2 and 3 in $T_1$ are idle and channel 1 in $T_2$ is idle. However, channel 1 in $T_2$ is sensed busy and it will not be allocated. Also, channel 2 in $T_2$ is incorrectly sensed to be idle and scheduling a request on this channel will lead to a collision. We let ${\pi}_1(i)$ denote the probability that channel $i$ in $T_1$ is idle and ${\pi}_2(j)$ the probability that channel $j$ in $T_2$ is idle. We also assume that the prior distribution of the PU activity is accurately acquired over time. We assume that state changes occur at the beginning of a time slot. Let $C\overset{\triangle}=|T_1|+|T_2|$ denote the total number of channels, which remains constant over time. Some of our technical results apply to the special case when all channels in
$T_2$ are homogenous, that is, when the channels have the same $\pi_2(i)$, $P_m(i)$ and $P_f(i)$. Thus, they also have the same $P_0(i)$ and $P_I (i)$.  

The availabilities of channels in $T_1$ and $T_2$ at $t$ are denoted by binary vectors $\vec{I}_1(t)=(I_1^1(t),\cdots,I_1^k(t),\cdots)$ and $\vec{I}_2(t)=(I_2^1(t),\cdots,I_2^l(t),\cdots)$, respectively, where $0$ represents idle and $1$ represents busy states. Moreover, $\vec{I}_2^s(t)$ denotes the sensed availabilities of channels in $T_2$ at $t$. \ignore{To differentiate the terms, the observed idleness for channels in $T_1$ means real idleness while the sensed idleness for channels in $T_2$ may due to sensing inaccuracies. We define function $Z(\cdot)$ as the number of $0$ elements in a vector. For instance, $Z(\vec{I}_1(t))$ indicates the number of idle channels in $T_1$ at time $t$.} Let $P_f(k)$, $k\in T_2$, denote the {\bf probability of false alarm} for channel $k$, i.e., the probability that SUs cooperatively sense channel $k$ to be busy given that it is actually idle. Let $P_m(k)$ represent the {\bf probability of misdetection} for channel $k$, i.e., the probability that SUs cooperatively sense channel $k$ to be idle given that it is actually busy.  We further define $P_I(k)$ as the probability that channel $k$ is sensed idle and $P_0(k)$ as the conditional probability of channel $k$ being idle given that it is sensed idle. Note that $P_I(k)={\pi}_2(k)(1-P_f(k))+(1-{\pi}_2(k))P_m(k)$ and $P_0(k)=\frac{{\pi}_2(k)(1-P_f(k))}{P_I(k)}$. We assume that $P_f(k)$ and $P_m(k)$ are constant for any channel $k\in T_2$, which occurs e.g. when SUs are static in the system.

We assume each spectrum request is for a single time-frequency chunk, i.e., a single time slot of any channel in $T_1$ or $T_2$. \ignore{We are interested in requests with delay tolerance in the system.} Each request $i$ submitted at time $t$ is of the form $(a_i,d_i,w_i)$, where $a_i \geq t$ is the required service starting time, $d_i$ is the deadline, and $w_i$ is the valuation of request $i$, which will be added to the social welfare if request $i$ is served by $d_i$. We denote the set of requests by $\mathcal{N}=\{1,\cdots,N\}$. $H=\max_{i\in \mathcal{N}}{d_i}-\min_{i\in \mathcal{N}}{a_i}$ denotes the time period spectrum allocation needs to be made, and $\min_{i\in \mathcal{N}}{a_i}$ is normalized to $1$. The maximum number of outstanding requests in the system at any time is denoted as $r$. Table~\ref{tab:not} summarizes the notations used in the paper.

\begin{table}[t]
\centering
\caption{{\scriptsize Notation List}}
\vspace{-1em}
\begin{tabular}{|c|l|}
\hline
{\scriptsize Symbol}&{\footnotesize Meaning}\\\hline\hline
{\scriptsize $\mathcal{N}$}&{\scriptsize Set of spectrum requests submitted to the operator}\\\hline
{\scriptsize $T_1$}&{\scriptsize Set of channels managed by the operator}\\\hline
{\scriptsize $T_2$}&{\scriptsize Set of channels not managed by the operator}\\\hline
{\scriptsize $\pi_1(i)$}&{\scriptsize Probability that channel $i$ in $T_1$ is idle}\\\hline
{\scriptsize $\pi_2(j)$}&{\scriptsize Probability that channel $j$ in $T_2$ is idle}\\\hline
{\scriptsize $C$}&{\scriptsize The total number of channels ($C=|T_1|+|T_2|$)}\\\hline
{\scriptsize $\vec{I}_1(t)$}&{\scriptsize Availabilities of channels in $T_1$ at $t$}\\\hline
{\scriptsize $\vec{I}_2(t)$}&{\scriptsize Availabilities of channels in $T_2$ at $t$}\\\hline
{\scriptsize $\vec{I}_2^s(t)$}&{\scriptsize Sensed availabilities of channels in $T_2$ at $t$}\\\hline
{\scriptsize $P_f(k)$}&{\scriptsize Probability of false alarm for channel $k\in T_2$}\\\hline
{\scriptsize $P_m(k)$}&{\scriptsize Probability of misdetection for channel $k\in T_2$}\\\hline
{\scriptsize $P_I(k)$}&{\scriptsize Probability that channel $k$ is sensed idle}\\\hline
{\scriptsize $P_0(k)$}&{\scriptsize Probability of channel $k$ being idle given that it is sensed idle}\\\hline
{\scriptsize $a_i$}&{\scriptsize Earliest service time for request $i$}\\\hline
{\scriptsize $d_i$}&{\scriptsize Deadline of request $i$}\\\hline
{\scriptsize $w_i$}&{\scriptsize Valuation of the request $i$}\\\hline
{\scriptsize $H$}&{\scriptsize The time period where spectrum allocation has to be made}\\\hline
{\scriptsize $r$}&{\scriptsize Maximum number of outstanding requests in the system at any time}\\\hline
{\scriptsize $Q$}&{\scriptsize Penalty price per collision}\\\hline
\end{tabular}
\label{tab:not}
\vspace{-1em}
\end{table}

We are interested in maximizing the social welfare of the operator and the SUs in the system: When an SU submits a request, the operator will not make a commitment on the service; the valuation $w_i$ is added to the social welfare if request $i$ is served by $d_i$. The social welfare is defined as the total valuations from the requests served minus the collision cost to channels in $T_2$. Let $Q$ denote the penalty incurred per collision. Let $x_{il}(t)$ ($i\in \mathcal{N}$, $l\in T_1\cup T_2$, $t=1,\cdots,H$) denote the allocation indicator: $x_{il}(t)=1$ if request $i$ is allocated to channel $l$ at $t$; $x_{il}(t)=0$ otherwise. $y_i$ denotes the service indicator: $y_i=1$ if request $i$ is served by $d_i$; $y_i=0$ otherwise. The social welfare maximization problem is then formulated as follows, where $Z(\cdot)$ denotes the number of $0$ elements in a vector:

Problem~(A):
\[\max\limits_{{\bf{x}}, \vec{y}}{\text{E}_{\bf{I_1}, \bf{I_2},\bf{I_2^s}}\Big[\sum\limits_{i\in \mathcal{N}}{y_i w_i}-Q\sum_{i\in \mathcal{N}}{\sum_{l\in T_2}{\sum_{t=1}^{H}{x_{il}(t)I_2^l(t)}}}\Big]}\]

\begin{equation}
\mbox{s.t. }\sum_{t=a_i}^{d_i}{\Big(\sum_{l\in T_1}{x_{il}(t)}+\sum_{k\in T_2}{x_{ik}(t)(1-I_2^k(t))}\Big)}\ge y_i,
\label{eq:serve}
\end{equation}
\[
\mbox{for all }i\in \mathcal{N}
\]
\begin{equation}
\sum_{i\in \mathcal{N}}{\sum_{l\in T_1}{x_{il}(t)}}\le Z(\vec{I}_1(t))\mbox{, for all $t=1,\cdots,H$}
\label{eq:alloc1}
\end{equation}
\begin{equation}
\sum_{i\in \mathcal{N}}{\sum_{k\in T_2}{x_{ik}(t)}}\le Z(\vec{I}_2^s(t))\mbox{, for all $t=1,\cdots,H$},
\label{eq:alloc2}
\end{equation}
\noindent where ${\bf{x}}=(x_{il}(t))_{i,l,t}$, $\vec{y}=(y_i)_{i\in \mathcal{N}}$, ${\bf{I_1}}=(\vec{I}_1(t))_{t=1,\cdots,H}$, ${\bf{I_2}}=(\vec{I}_2(t))_{t=1,\cdots,H}$, ${\bf{I_2^s}}=(\vec{I}_2^s(t))_{t=1,\cdots,H}$. The cost $Q\sum_{i\in \mathcal{N}}{\sum_{l\in T_2}{\sum_{t=1}^{H}{x_{il}(t)I_2^l(t)}}}$ takes into account the current availabilities of channels in $T_2$. Inequality~(\ref{eq:serve}) reflects the relationship between the allocation indicator $x_{il}(t)$ and the service indicator $y_i$. Inequality~(\ref{eq:alloc1}) guarantees that a channel in $T_1$ will not be allocated unless it is observed idle. Likewise, Inequality~(\ref{eq:alloc2}) guarantees that a channel in $T_2$ will not be allocated unless it is sensed idle.

The challenges of Problem~(A) are threefold: 1) The requests are uncertain since they may be submitted at different time slots; 2) Spectrum availabilities of $T_1$ and $T_2$ in the future are not known at the current time slot; 3) Sensing is not accurate for channels in $T_2$. In the following, we propose an offline optimal solution in Section~\ref{sec:off_wo} and an online solution in Section~\ref{sec:on_wo}. We define the offline algorithm as an algorithm that decides the channel allocation for outstanding requests in each time slot with only the observed availabilities of channels in $T_1$ and sensed availabilities in $T_2$ of the current slot. All requests, including future arrivals, are assumed to be known. For instance, SUs submit their requests at the beginning of $H$. The operator then knows the full arrival information. In each time slot, the operator has to make channel allocation decisions based on the observed availabilities of its own channels and the sensed availabilities of channels managed by other operators. The only difference between online and offline algorithms is that online algorithm does not assume the full arrival information to be known ahead of time. {\it Both algorithms are designed under the challenges of spectrum uncertainty and sensing inaccuracy.} 
\section{Optimal Offline Algorithm}
\label{sec:off_wo}
In this section, we study Problem~(A) under the assumption that the operator has full knowledge of the spectrum requests in advance. By our assumptions on channel statistics, the problem can be modeled as a finite horizon Markov Decision Process (MDP) \cite{Puterman}. In this section, we propose an optimal dynamic programming based solution to the problem. We start with the simple case where $T_1 = \emptyset$ and all the channels for serving SUs are in $T_2$, which models the case where all the channels owned by the operator are overloaded by PU traffic. Then, we proceed with the general case where both $T_1$ and $T_2$ channels are available for use in the system. In each time slot, based on the knowledge of the spectrum requests and the current channel state, the operator makes a joint decision including 1) which subset of requests to schedule; 2) which subset of channels to allocate; 3) which request to assign to which channel. In our solution, we consider all possible scenarios for each time slot and find the schedule that maximizes the expected social welfare. Note that the social welfare is composed of two parts: Valuations of SUs that are served and the cost caused by collisions on the channels in $T_2$. We show that our algorithm has a complexity of $O(2^r 3^C (\max{\{C,r\}})^{\min{\{C,r\}}} HCr)$. When $(a_i,d_i)$ of requests do not have a dense overlap, i.e., $r=O(\log N)$ where $N$ is the total number of requests in $[1,H]$, our algorithms are of polynomial complexity.

We first define $F(D, t)$ as the maximum expected social welfare from the beginning of slot $t$ till the end of slot $H$ given that the set of outstanding requests is $D$. The expectation takes into account all possible channel realizations and sensing results. We define $F(D,H+1)=0$ for all $D$. Our goal is to calculate $F(D,1)$ where $D=\{i: i\in \mathcal{N}, a_i=1\}$ (Algorithm~\ref{alg:off}). We calculate it backward from $t=H$ till $t=1$ is reached since requests requiring service in future time slots have an impact on the current optimal scheduling decision. Note that at any time $t$, it is sufficient to consider $D$ in $F(D, t)$'s for being any subset of the requests that satisfy $a_i\le t < d_i$. \ignore{For instance, a request not scheduled in the current slot may compete for channel resources with future requests.}


\begin{algorithm*}[t]
    \caption{Dynamic Programming based Optimal Algorithm for Social Welfare Maximization}\label{alg:off}
    {\bf Offline computation}
	\begin{algorithmic}[1]

	\FOR{$t=H$ to $1$}
		\FORALL{$D\subseteq \{i:a_i\le t<d_i\}$}
			\FORALL{$S\subseteq T_2$}
				\STATE $X(D,S, t)\leftarrow  \max\limits_{{\bf x}(t)}{\Bigg[\sum\limits_{S_1 \subseteq S}{\prod\limits_{l\in S_1}{P_0(l)\prod\limits_{m\in S\setminus S_1}{(1-P_0(m)) (\sum\limits_{n\in D}{\Big[ w_n \sum\limits_{k\in S_1}{x_{nk}(t) }-Q \sum\limits_{k\in S\setminus S_1}{x_{nk}(t)  } \Big]}+F(D',t+1))}}}\Bigg]}$
			\ENDFOR
			\STATE $F(D,t)\leftarrow\sum\limits_{S \subseteq T_2}{\prod\limits_{l\in S}{P_I(l)\prod\limits_{m\in T_2\setminus S}{(1-P_I(m))}X(D, S,t)}}$
		\ENDFOR
	\ENDFOR
	\end{algorithmic}
	\vspace{2ex}{\bf Real-time scheduling}
	\begin{algorithmic}[1]
	\STATE  At each time slot $t$ with a set of requests $D$ that are currently in the system and a set of channels $S$ that are sensed idle, allocate channels to the requests based on the schedule ${\bf x}(t)$ that maximizes $X(D,S, t)$.
	\end{algorithmic}
    \vspace{-0.2em}
\end{algorithm*}

\subsection{With no available channels in $T_1$}

When no channel is in $T_1$, the spectrum bands managed by the operator, SUs can only be served by channels in $T_2$. SUs may request spectrum in arbitrary time slots. The success of serving request $i$ contributes $w_i$ to the social welfare while the assignment of a request to a busy channel causes collisions, incurring a penalty of $Q$.

We define $X(D, S, t)$ as the maximum expected social welfare from $t$ ($t=1,\cdots,H$) to the end of the period, given that the set of outstanding requests is $D$ and channels in $S$ are sensed idle ($S\subseteq T_2$). The expectation is taken over all possible realizations of ${\bf I_2}$. Then, 
\begin{eqnarray}
X(D, S, t)=  &&\max\limits_{{\bf x}(t)}{\Bigg[\sum\limits_{S_1 \subseteq S}{\prod\limits_{l\in S_1}{P_0(l)\prod\limits_{m\in S\setminus S_1}{(1-P_0(m))}}}}\nonumber\\
&&( W(D, S, S_1,{\bf {x}}(t), t)+F(D',t+1))\Bigg],
\label{eq:rec}
\end{eqnarray}

\noindent where $W(D, S, S_1,{\bf {x}}(t), t)$
\[= \sum\limits_{n\in D}{\Big[ w_n \sum\limits_{k\in S_1}{x_{nk}(t) }-Q \sum\limits_{k\in S\setminus S_1}{x_{nk}(t)  } \Big]}\]
\noindent is defined as the social welfare achieved in time slot $t$, for a given $D$, the set of outstanding requests; $S\subseteq T_2$, the set of channels sensed idle; $S_1\subseteq S$, the set of channels that are sensed idle and actually idle; and ${\bf {x}}(t)$, the channel allocation at $t$. Recall that $x_{nk}(t)$ is the allocation indicator used to determine whether the SU is served by this allocation. We form $D'$ based on $D$ as follows: If request $m$ is allocated to channels in $S_1$, then remove $m$ from $D$, which means it is served and the request no long exists. If request $n$ satisfies $a_n=t+1$, then add $n$ to $D$, which indicates it is a new request. Among the remaining requests, those that expire at the beginning of $t+1$ are removed from $D$.

Based on $X(D,S, t)$, we calculate $F(D,t)$ as follows. The expectation in $F(D, t)$ in the form of the product of $P_I(l)$ and $(1-P_I(m))$ takes into account all realizations of ${\bf I_2^s}$.
\begin{equation}
F(D, t)= \sum\limits_{S \subseteq T_2}{\prod\limits_{l\in S}{P_I(l)\prod\limits_{m\in T_2\setminus S}{(1-P_I(m))}X(D, S,t)}},
\label{eq:F}
\end{equation}

In Algorithm~\ref{alg:off}, our objective $F(\{i: i\in \mathcal{N}, a_i=1\},1)$ is calculated by dynamic programming. It first calculates the maximum social welfare and the corresponding schedule for each time slot, and then specifies the real time operations. Lines~1-6 calculate $F(D,t)$ backward from $H$ to $1$ given the initial condition defined earlier $F(D,H+1)=0$ for all $D$. Line~5 calculates the optimal scheduling policy for time $t$ given $D$, the request set; $S$, the set of channels sensed idle; and $S_1$, the set of channels sensed idle and actually idle, according to Equation~(\ref{eq:rec}). The value of $F(D,t)$ is updated in Line~6 according to Equation~(\ref{eq:F}). The complexity of the Equation~(\ref{eq:F}) is $O(3^{|T_2|} (\max{\{|T_2|,r\}})^{\min{\{|T_2|,r\}}} |T_2| r)$: The number of possible channels realizations is $3^{|T_2|}$ since different social welfare values will be generated in the cases where the channel is sensed idle but actually busy, it is sensed idle and actually idle, and all other cases. It takes at most $(\max{\{|T_2|,r\}})^{\min{\{|T_2|,r\}}}$ combinations to find the optimal ${\bf x}$ in Equation~(\ref{eq:rec}). The complexity for the calculation of $W(D, S, S_1,{\bf {x}}(t), t$ is $O(|T_2| r)$. On the other hand, given $t$, the number of possible argument combinations in $F(D,t)$ is $O(2^r H)$ by assumption. The total time complexity is $O(2^r 3^C (\max{\{C,r\}})^{\min{\{C,r\}}} HCr)$. Note that $C$ is assumed to be a constant in our model.

\subsection{With at least one channel in $T_1$}

With channels in $T_1$, requests can be served by channels in both $T_1$ and $T_2$. Since the channel availabilities of $T_1$ are known at the beginning of each time slot, they can serve SU requests without any cost. Thus, once observed idle, channels in $T_1$ could be assigned to requests so as to maximize the sum of valuations. Our focus is still the allocation of channels in $T_2$ if they are sensed idle.

We define $Y(D, S,t)$ as the maximum expected social welfare from $t$ to the end of the period, given that the set of outstanding requests is $D$ and channels in $S$ are sensed idle ($S\subseteq T_2$). We also define $\hat{X}(D,\Gamma, S, t)$ as the maximum expected social welfare from $t$ to the end of the period, given that the set of outstanding requests is $D$, channels in $\Gamma$ are observed to be idle ($\Gamma \subseteq T_1$), and channels in $S$ are sensed idle ($S \subseteq T_2$). The expectation in $Y(D,S,t)$ is for all realizations of ${\bf I_1}$ and ${\bf I_2}$. The expectation in $\hat{X}(D,\Gamma, S, t)$ is for all realizations of ${\bf I_2}$. Then,
\[\hat{X}(D,\Gamma,S, t)=  \max\limits_{{\bf x}(t)}{\Bigg[\sum\limits_{S_1 \subseteq S}{\prod\limits_{l\in S_1}{P_0(l)\prod\limits_{m\in S\setminus S_1}{(1-P_0(m))}}}}\]
\begin{equation}
( \hat{W}(D,\Gamma, S, S_1, {\bf x}(t),t)+F(D',t+1))\Bigg],
\label{eq:rec2}
\end{equation}
\noindent where
\[\hat{W}(D,\Gamma, S, S_1, {\bf x}(t),t)=\sum\limits_{n\in D}{\Big[ w_n (\sum\limits_{l\in \Gamma}{x_{nl}(t)  }+\sum\limits_{k\in S_1}{x_{nk}(t)  })}\]
\[-Q \sum\limits_{k\in S\setminus S_1}{x_{nk}(t) } \Big]\]
\noindent is defined as the social welfare achieved in time slot $t$, given $D$, the set of outstanding requests; $\Gamma$, the set of channels in $T_1$ that are observed to be idle; $S$, the set of channels sensed idle; $S_1$, the set of channels sensed idle and actually idle; and ${\bf x}(t)$, the channel allocation at $t$. The only difference between $W(D, S, S_1,{\bf {x}}(t), t)$ and $\hat{W}(D,\Gamma, S, S_1, {\bf x}(t),t)$ is the addition of the valuations contributed by the service on channels in $T_1$. We form $D'$ based on $D$ in a similar way to Equation~(\ref{eq:rec}): If a request $m$ is allocated to channels in $S_1\cup \Gamma$, then remove $m$ from $D$, which means it is served and the request no long exists. If request $n$ satisfies $a_n=t+1$, then add $n$ into $D$, which indicates it is a new request. All other SU $i$ are removed from $D$ only when $d_i=t+1$. We then calculate $Y(D,S, t)$ as:
\begin{equation}
Y(D,S, t)=\sum\limits_{\Gamma\subseteq T_1}{\prod\limits_{l\in \Gamma}{\pi_1(l)\prod\limits_{m\in T_1\setminus \Gamma}{(1-\pi_1(m))}\hat{X}(D,\Gamma, S,t)}}.
\label{eq:Y}
\end{equation}

Hence,
\begin{equation}
F(D, t)= \sum\limits_{S \subseteq T_2}{\prod\limits_{l\in S}{P_I(l)\prod\limits_{m\in T_2\setminus S}{(1-P_I(m))}Y(D,S,t)}},
\label{eq:F2}
\end{equation}

\noindent which takes into account all realizations of ${\bf I_1}$, the availabilities of channels in $T_1$; ${\bf I_2^s}$, the sensed availabilities of channels in $T_2$; and ${\bf I_2}$, the actual availabilities of channels in $T_2$.

The algorithm is similar to Algorithm~\ref{alg:off} except that $F(D,t)$ is updated according to Equation~(\ref{eq:F2}). The complexity of Equation~(\ref{eq:F2}) is $O(3^{|T_2|} 2^{|T_1|} (\max{\{|T_2|,r\}})^{\min{\{|T_2|,r\}}} C r)$. The difference from the complexity of Equation~(\ref{eq:F}) lies in $2^{|T_1|}$, which is caused by the number of $T_1$ channel realizations. Following a similar argument as in the case where $|T_1|=0$, the total time complexity is still $O(2^r 3^C (\max{\{C,r\}})^{\min{\{C,r\}}} HCr)$. Note that, for homogeneous channels in $T_2$, the allocation policy becomes simpler since allocation to different channels in $T_2$ makes no difference. Then, we can replace $(\max{\{C,r\}})^{\min{\{C,r\}}}$ with $r$ in total complexity, resulting in a complexity of $O(2^r 3^C HCr^2)$.

\subsection{Discussion}
In this section, we prove some structural properties of the optimal solution, which helps to further reduce the time complexity of the algorithm and also provides insight to the design of the online algorithm discussed in Section~\ref{sec:on_wo}. Note that at any time $t$, for an active request $i$ and a channel $k \in T_2$ that is sensed idle, $P_0(k) w_i - Q (1-P_0(k))$ is the expected immediate social welfare contributed by request $i$ if $i$ is assigned to $k$ in the current slot. Proposition~\ref{prop:nec} shows that a non-negative immediate social welfare is necessary for request $i$ to be served by channel $k$ in the optimal solution, which turns out to be a sufficient condition in certain scenario as stated in Proposition~\ref{prop:suf}, as well. \ignore{Proposition~\ref{prop:cost} shows the expected cost per service on each channel and suggests sorting channels by this cost to further reduce the channel allocation complexity.}

\begin{proposition}
At any time $t$, if a request $i$ is scheduled on channel $k \in T_2$ in Algorithm~\ref{alg:off}, then $P_0(k) w_i \ge Q (1-P_0(k))$.
\label{prop:nec}
\end{proposition}

\begin{proof}
We will prove Proposition~\ref{prop:nec} for the case without $T_1$ channels. It can be shown for the general case in a similar way. Suppose at time $t$, request $i$ is assigned to channel $k$ in the optimal solution, with the system state being $(D, S, S_1)$ as defined before. Note that $k$ may or may not be in $S_1$. Let ${\bf x}(t)$ be the optimal schedule, $D_S$ be the set of requests scheduled in ${\bf x}(t)$, and $D'_{S,S_1}$ be the set of outstanding requests for $t+1$ with $D_S$ scheduled in $t$. Let ${\bf \hat{x}}(t)$ be the same schedule as ${\bf x}(t)$ except that $i$ is excluded. To simplify the notation, let $R(S,S_1)=\prod\limits_{l\in S_1}{P_0(l)\prod\limits_{m\in S\setminus S_1}{(1-P_0(m))}}$. Then we have

 \[X(D,S,t) = \Bigg[\sum\limits_{S_1 \subseteq S, k\in S_1}{R(S,S_1)(W(D,S,S_1,{\bf \hat{x}}(t),t) } \]
 \[+w_i+F(D'_{S,S_1},t+1)) \Bigg]\]
 \[+ \Bigg[\sum\limits_{S_1 \subseteq S, k\notin S_1}{R(S,S_1)(W(D,S,S_1,{\bf \hat{x}}(t),t) } \]
 \begin{equation}
-Q+F(D'_{S,S_1},t+1)) \Bigg]
\label{eq:DS}
\end{equation}

On the other hand, if $i$ is not scheduled, then the expected social welfare from $t$ to $H$ is
 \[X'(D,S,t) = \Bigg[\sum\limits_{S_1 \subseteq S, k\in S_1}{R(S,S_1)(W(D,S,S_1,{\bf \hat{x}}(t),t) } \]
 \[+F(D'_{S,S_1}\cup \{i\},t+1)) \Bigg]\]
 \[= \Bigg[\sum\limits_{S_1 \subseteq S, k\notin S_1}{R(S,S_1)(W(D,S,S_1,{\bf \hat{x}}(t),t) } \]
\begin{equation}
+F(D'_{S,S_1},t+1)) \Bigg]
\label{eq:DSnoi}
\end{equation}

Since $D_S$ is the optimal solution for Equation~(\ref{eq:rec}), we have (\ref{eq:DS})-(\ref{eq:DSnoi})$\ge 0$. By rearranging the terms, we obtain
\[\sum\limits_{S_1 \subseteq S, k\in S_1}{R(S,S_1) (F(D'_{S,S_1},t+1)-F(D'_{S,S_1}\cup \{i\},t+1))}  \]
\[+w_i \sum\limits_{S_1 \subseteq S, k\in S_1}{R(S,S_1)}-Q \sum\limits_{S_1 \subseteq S, k\notin S_1}{R(S,S_1)}\]
\[=\sum\limits_{S_1 \subseteq S, k\in S_1}{R(S,S_1) (F(D'_{S,S_1},t+1)-F(D'_{S,S_1}\cup \{i\},t+1))} \]
\begin{equation}
+(w_i P_0(k)-Q (1-P_0(k)))\sum\limits_{S_1 \subseteq S\setminus \{k\}}{R(S,S_1)}\ge 0,
\label{eq:diff}
\end{equation}

\noindent where $F(D'_{S,S_1},t+1)-F(D'_{S,S_1}\cup \{i\},t+1)\le 0$ since the social welfare is monotonic over the set of requests. Hence, we must have $P_0(k) w_i \ge Q (1-P_0(k))$ for Inequality~(\ref{eq:diff}) to hold.
\end{proof}

Proposition~\ref{prop:suf} shows that the condition $P_0(k) w_i >Q (1-P_0(k))$ is also sufficient for a request to be scheduled for homogenous channels. To simplify notation, we drop the index for channel related parameters for the homogeneous case.

\begin{proposition}
In a system with no channels in $T_1$ and homogeneous channels in $T_2$, if there exists at least one request $i$ that satisfies $P_0 w_i>Q (1-P_0)$ in a slot $t$ and there is at least one channel sensed idle, then in the optimal solution at least one of the requests satisfying this condition will be scheduled, for all $t$.
\label{prop:suf}
\end{proposition}
\begin{proof}
Given the system state $(D, S, S_1)$ at time $t$, consider a subset of requests $D_S\subseteq D$ to be scheduled where $i \in D_S$. Let $D'_{S,S_1}$ denote the set of the outstanding requests at $t+1$ given that $D_S$ is scheduled at $t$, and $k$ the channel assigned to $i$ in the schedule. We want to show that the expected social welfare from $t$ to the end of the time period with $D_S$ scheduled at $t$ is at least as large as that with $D_S\setminus \{i\}$ scheduled. We define $U(T_2,S)=\prod\limits_{l\in S}{P_I(l)\prod\limits_{m\in T_2\setminus S}{(1-P_I(m))}}$. We also define $F_1(D,t)$ as the expected social welfare from $t$ till the end of $H$ by scheduling $D_S$ at time $t$ and $F_2(D,t)$ as the expected social welfare from $t$ till the end of $H$ by scheduling $D_S\setminus \{i\}$ in time $t$. By Equations~(\ref{eq:DS}), (\ref{eq:DSnoi}), and (\ref{eq:diff}) we obtain
\[ F_1(D,t)-F_2(D,t) =\sum\limits_{S \subseteq T_2}{U(T_2,S)}\Bigg[\sum\limits_{S_1 \subseteq S, k\in S_1}{P_0^{|S_1|}(1-P_0)^{|S\setminus S_1|}}\]
\[ \Big[F(D'_{S,S_1},t+1)-F(D'_{S,S_1}\cup \{i\},t+1)\Big]  \]
\[+(w_i P_0-Q (1-P_0))\sum\limits_{S_1 \subseteq S\setminus \{k\}}{P_0^{|S_1|} (1-P_0)^{|S\setminus S_1|-1}}\Bigg] \]
\[=\sum\limits_{S \subseteq T_2}{U(T_2,S) \Big[ P_0 (F(D'_{S,S_1},t+1)-F(D'_{S,S_1}\cup \{i\},t+1))}\]
\begin{equation}
+(w_i P_0 -Q (1-P_0))\Big] \sum\limits_{S_1 \subseteq S\setminus \{k\}}{P_0^{|S_1|} (1-P_0)^{|S\setminus S_1|-1}}
\label{eq:diff2}
\end{equation}

In the following we will show that $P_0(F(D'_{S,S_1}\cup \{i\},t+1)- F(D'_{S,S_1},t+1))\le w_i P_0 -Q (1-P_0)$. We first observe that $F(D,t)\le F(D\setminus \{i\},t)+F(\{i\},t)$ for any $t$, $D$ and $i\in D$ since $i$ is competing with requests in $D\setminus \{i\}$ for the spectrum in the former case but not in the latter case. in \ignore{We observe that the maximum social welfare achieved from $t$ to the end of $H$ with the outstanding request set $D$ at $t$ is no more than that achieved by splitting $D$ into two sets $D\setminus \{i\}$ and $\{i\}$ at $t$ while letting each one of them taking the same amount of resource given to $D$. By submodularity,} Hence we only need to prove that $P_0(F(\{i\},t+1)-F(\emptyset,t+1))\le w_i P_0 -Q (1-P_0)$ for all $t$, that is, $P_0 F(\{i\},t+1)\le w_i P_0 -Q (1-P_0)$ for all $t$ since $F(\emptyset,t+1)=0$. We will prove it by induction. We define $\hat{P}_I$ as the probability that at least one channel is sensed idle. We start with $t=H$ and then $P_0 F(\{i\},H)=P_0 \hat{P}_I(w_i P_0 -Q (1-P_0))\le w_i P_0 -Q (1-P_0)$. Suppose $P_0 F(\{i\},t)\le w_i P_0 -Q (1-P_0)$ for all $t> \tau$, then $F(\{i\},\tau)=(1-\hat{P}_I)F(\{i\},\tau+1)+\hat{P}_I \max{\{w_i P_0 + (1-P_0) (F(\{i\},\tau+1)-Q),F(\{i\},\tau+1)\}}$. We calculate
\[F(\{i\},\tau+1)-(w_i P_0 + (1-P_0) (F(\{i\},\tau+1)-Q))\]
\[=P_0(F(\{i\},\tau+1)-(w_i P_0 -Q (1-P_0)))\overset{(a)}\le 0,\]

\noindent where (a) is by the induction assumption. Then we know $i$ should be scheduled in the optimal solution at $\tau$ if it is the only request. Hence,
\[P_0 F(\{i\},\tau)=P_0((1-\hat{P}_I)F(\{i\},\tau+1)\]
\[+\hat{P}_I(w_i P_0 + (1-P_0) (F(\{i\},\tau+1)-Q)))\]
\[=(1-P_0\hat{P}_I)P_0 F(\{i\},\tau+1)+ P_0\hat{P}_I (w_i P_0 -Q (1-P_0)))\]
\[\overset{(b)}\le w_i P_0 -Q (1-P_0)),\]
\noindent where (b) is by the induction assumption. Thus $P_0(F(D\cup \{i\},t)-F(D,t))\le w_i P_0 -Q (1-P_0)$ for all $D$ and $t$. Therefore, Equation~(\ref{eq:diff2})$\ge 0$, which means the expected social welfare from $t$ to the end of the time period with $D_S$ scheduled at $t$ is always better than that with $D_S\setminus \{i\}$ scheduled.
\end{proof}

Based on these propositions, we can reduce the candidate set of requests for scheduling in each time slot. For instance, no requests should be scheduled if $P_0(k) w_i\le Q (1-P_0(k))$ for all existing requests $i$ and all $k$ sensed idle. Also, in a system with no channels in $T_1$ and homogeneous channels in $T_2$, the candidate set is composed of all requests that satisfy $P_0(k) w_i>Q (1-P_0(k))$. We utilize these propositions in the design of our online algorithm.

\ignore{
On the other hand, we can first sort the channels by $c_k\overset{\triangle}=Q(1-P_0(k))/P_0(k)$ in ascending order before collecting the sensing results. In Line~4 of Algorithm~\ref{alg:off}, channels are searched in the same order to calculate the optimal schedule ${\bf x}(t)$. The complexity of the optimal schedule selection is then reduced to $O(C\log{C}+r)$. We have the following observation for any algorithm in Proposition~\ref{prop:cost}, which will also be used in Sections~\ref{sec:on_wo} and \ref{sec:inc}.

\begin{proposition}
\label{prop:cost}
For any algorithm, $c_k$ is the expected cost per a request service on channel $k$.
\end{proposition}

\begin{proof}
For any algorithm, consider the time interval right after a request is served by channel $k$ and before the next request is served by channel $k$. Remove all time slots in the interval when there are no requests in the system or channel $k$ is sensed but not allocated. Given that a channel is sensed idle, the probability that collision happens is $1-P_0(k)$. Thus the number of slots where collisions happen follows a geometric distribution and the expected cost per a request service on channel $k$ is $Q(1-P_0(k))/P_0(k)$.
\end{proof}}

\section{Online Algorithm}
\label{sec:on_wo}
In this section, we introduce a greedy online algorithm (Algorithm~\ref{alg:on}) that does not need future arrival information. For systems where requests are not submitted ahead of the required service starting time $a_i$, the online algorithm makes decisions based on the information available in the current slot.

\begin{algorithm}[t]
    \caption{Greedy Online Algorithm}\label{alg:on}
    In each time slot $t$:
    \begin{algorithmic}[1]
	\IF{$D=\emptyset$}
		\STATE exit
	\ENDIF
	\STATE Sort channels in $S$ (channels sensed idle in $T_2$) by $c_k$ in ascending order
	\STATE Sort requests in $D$ (outstanding ones) by $w_j$ in descending order
	\STATE $i\leftarrow 1$
	\FORALL{$l$ in $\Gamma$ (channels in $T_1$ observed idle)}
		\STATE $x_{il}(t)\leftarrow 1$
		\STATE $D\leftarrow D\setminus \{i\}$
		\IF{$D=\emptyset$}
			\STATE break
		\ENDIF
		\STATE $i\leftarrow i+1$
	\ENDFOR
	\IF{$D=\emptyset$}
		\STATE exit
	\ENDIF
	\STATE $n\leftarrow |\Gamma|+1$
	\FORALL{$k$ in $S$ (channels sensed idle in $T_2$)}
		\IF{$w_n \le \theta(k)$}
			\STATE break
		\ENDIF
		\STATE $x_{nk}\leftarrow 1$
		\STATE $D\leftarrow D\setminus \{n\}$
		\IF{$D=\emptyset$}
			\STATE break
		\ENDIF
		\STATE $n\leftarrow n+1$
	\ENDFOR
    \end{algorithmic}
    \vspace{-0.2em}
\end{algorithm}

In Algorithm~\ref{alg:on}, the main idea is to (greedily) offer requests with higher valuation channels with better quality. We define $c_k\overset{\triangle}=Q(1-P_0(k))/P_0(k)$, which is the expected cost of serving one request on channel $k$ (will be shown in Lemma~\ref{lem:cost}). Note that $c_k = 0$ for $k \in T_1$. Lines~3 and 4 sort channels sensed idle by $c_k$ and current requests by $w_j$, respectively. Since accessing channels in $T_1$ causes no cost if observed idle, they are allocated first to requests with highest valuations (Lines~6-11). In Lines~15-22, the remaining requests are allocated to channels in $T_2$ sensed idle from highest valuation to lowest if they satisfy $w_n > \theta(k)$ where $\theta(k)$ serves as a {\it reservation price} for using channel $k$. We set $\theta (k) = c_k$ in this section, which is motivated by Propositions~\ref{prop:nec} and \ref{prop:suf}. Allowing different values of reservation price provides a way for trading off the social welfare and the revenue of the operator, which will be discussed in detail in Section~\ref{sec:inc}.

The time complexity of Algorithm~\ref{alg:on} is $O(C\log{C}+r\log{r})$ since the complexity of sorting in Lines~3 and 4 dominates that of allocation in Lines~5-22. We then study the performance of the online algorithm. An online algorithm for a maximization problem is $c$-competitive ($c \leq 1$) if it achieves at least a fraction $c$ of the objective value of an optimal offline algorithm for any finite input sequence~\cite{competitive-analysis}. We show that the greedy online algorithm is 1/2-competitive when $|T_1| = 0$ and channels in $T_2$ are homogeneous in Proposition~\ref{prop:ratio}. For heterogenous channels, we will show that the online algorithm achieves performance comparable to the optimal offline algorithm by numerical results in Section~\ref{sec:simu}.
\ignore{Since we will consider a system with only homogeneous channels in $T_2$, we drop the index for channel related notations.}To establish Proposition~\ref{prop:ratio}, we first show that $c_k$ is the expected cost per a request served by channel $k$ in Lemma~\ref{lem:cost}.

\begin{lemma}
\label{lem:cost}
For any scheduling policy, $c_k$ is the expected cost of serving a request on channel $k$.
\end{lemma}

\begin{proof}
For any scheduling policy, consider the time interval right after a request is served by channel $k$ and before the next request is served by channel $k$. Remove all time slots in the interval when there are no requests in the system or channel $k$ is sensed but not allocated. Given that a channel is sensed idle, the probability that collision happens is $1-P_0(k)$. Thus the number of slots where collisions happen follows a geometric distribution and the expected cost per a request service on channel $k$ is $Q(1-P_0(k))/P_0(k)$.
\end{proof}

Based on Lemma~\ref{lem:cost}, we show the competitive ratio of Algorithm~\ref{alg:on} in special cases.

\begin{proposition}
\label{prop:ratio}
If $|T_1|=0$, and channels in $T_2$ are homogeneous, then Algorithm~\ref{alg:on} is 1/2-competitive.
\end{proposition}
\begin{proof}
Let the random variable $\gamma$ denote the set of requests that are eventually served by the algorithm. Let $P_0 = P_0(k)$ for any channel $k \in T_2$. Since the channels in $T_2$ are homogeneous, we have $c=Q(1-P_0)/P_0$, which is the expected cost for serving a single request by Lemma~\ref{lem:cost}. Then the expected social welfare can be written as follows:
\[
\sum\limits_{k=1}^{N}{\Big[ (\sum\limits_{|\gamma|=k}{\Pr(\gamma)}\sum\limits_{i\in \gamma}{w_i})-kc \Pr(|\gamma|=k)\Big]}
\]
\[=\sum\limits_{k=1}^{N}{\Big[ \sum\limits_{|\gamma|=k}{\Pr(\gamma)\sum\limits_{i\in \gamma}{(w_i-c)}}\Big]}\]

Note that the greedy algorithm always chooses the active request with highest valuation. For any sample path, consider the set of requests served by the optimal offline algorithm and those by the greedy algorithm with $w'_i=w_i-c$ as the valuation. We follow the same argument as in \cite{haj}: We consider any request $i$ that is scheduled offline but not online. Since request $i$ is not scheduled online, it is present at time $t$ and the greedy algorithm schedules another request $j$ in that slot, the valuation of request $j$ should be as least as large as that of request $i$. For any request $i$ that is allocated offline and also online, it makes the same contribution to the social welfare. Then the offline solution achieves a social welfare at most twice that in the online solution since $\frac{w'_j}{w'_i+w'_j}\ge \frac{1}{2}$. Therefore, Algorithm~\ref{alg:on} is 1/2-competitive.
\end{proof}


Note that the factor 2 in Proposition~\ref{prop:ratio} does not depend on request arrival patterns or channel related parameters. Algorithm~\ref{alg:on} can always achieve at least $\frac{1}{2}$ of the social welfare of the optimal offline algorithm (Algorithm~\ref{alg:off}) when the system is only composed of homogeneous $T_2$ channels.
\section{Achieving Incentive Compatibility}
\label{sec:inc}
In this section, we design an online auction scheme which utilizes the online greedy algorithm (Algorithm~\ref{alg:on}) together with a payment scheme to achieve incentive compatibility for SUs. Due to the collision penalty, however, a social welfare optimal auction may end up with a negative revenue for the operator, which is not reasonable since the operator may choose not to start the auction in the first place. We introduce a reservation price for resolving this problem, which also provides a way of trading off the social welfare and revenue.

\subsection{Incentive Compatibility for SUs}
\ignore{To define incentive compatibility, we need the following definitions. We first define $p_i$ as the payment of SU for having its request $i$ served. Then the \emph{net utility} for request $i$ is defined as: $u_i=w_i-p_i$ if request $i$ is served and $u_i=0$ if not. At $t=a_i$, the bid of request $i$ is submitted and the private information include $a_i$, the required service starting time; $d_i$, the deadline; and $w_i$, the valuation. The mechanism is said to be incentive compatible if all of the requests achieve the best utility when they truthfully reveal any private information asked for by the mechanism \cite{Nisan}. We design Auction~1 based on Algorithm~\ref{alg:on}, and show it satisfies incentive compatibility for SUs in Proposition~\ref{prop:inc}. The \emph{Critical price} is the reported valuation of a request under which it will not be scheduled in the auction with the valuations of other requests unchanged. We define $\hat{a}_i$ and $\hat{d}_i$ as the reported required service starting time and the deadline, respectively. We define early-arrival misreport as the request reporting $\hat{a}_i<a_i$ in the bid and likewise late-departure misreport as the request reporting $\hat{d}_i>d_i$ in the bid. In practice, both of them can easily be detected since the request is no longer in the system when either misreport occurs.}

When the available spectrum resource cannot satisfy all the requests, which is often the case, a selfish SU may choose to cheat on its valuation or arrival and deadline times to obtain some priority of being served. Such strategic behavior leads to a less efficient system. In this section, an online auction scheme is presented (see Auction~1) to suppress the cheating behavior. At any time slot $t$, the operator accepts bids of the form $(\hat{a}_i, \hat{d}_i, \hat{w}_i$), where $\hat{a}_i = t$ and $\hat{d}_i$ denote the reported required service starting time and the deadline, respectively, and $\hat{w}_i$ denotes the reported valuation. All these values could be different from the true values of request $i$. We assume there is no early-arrival misreport and late-departure misreport in the system, that is, $\hat{a}_i \ge a_i$ and $\hat{d}_i \le d_i$ in any bid. In practice, both of them can easily be detected since the request is no longer in the system when either misreport occurs.

Let $p_i$ denote the payment that the operator charges a SU for having its request $i$ served. The \emph{net utility} for request $i$ is defined as: $u_i=w_i-p_i$ if request $i$ is served and $u_i=0$ if not. A mechanism is said to be {\it dominant-strategy incentive compatible} (DSIC) if for any given sample path of channel state realizations and sensing realizations and a set of requests, each request maximizes its utility when it truthfully reveals the private information independent of the bids from other requests (adapted from Definition 16.5 in \cite{Nisan}). In Auction~1, for every request that is successfully served by its deadline, a \emph{critical price} is charged, which is defined as the maximum reported valuation under which it will not be served assuming the other bids are fixed.

\noindent {\bf Auction 1}: Requests $(\hat{a}_i, \hat{d}_i,\hat{w}_i)$ are reported to the operator at time $t= \hat{a}_i$.

\noindent {\bf (i)} At the beginning of each $t$, allocate requests according to Algorithm~\ref{alg:on}.

\noindent {\bf (ii)} Every request successfully served pays its critical price, collected at its reported deadline.

\begin{proposition}
\label{prop:inc}
Auction~1 is DSIC with no early-arrival and no late-departure misreports.
\end{proposition}

\begin{proof}
According to Theorem 16.13 in~\cite{Nisan}, to show that Auction~1 is DISC, it is sufficient to show that the mechanism is monotonic in terms of both valuation and timing. That is, for a given sample path of channel realizations and sensing realizations and a set of requests, if request $i$ submitting a bid $(\hat{a}_i,\hat{d}_i,\hat{w}_i)$ wins, then it continues to win if it instead submits a bid $(\hat{a}'_i,\hat{d}'_i,\hat{w}'_i)$ with $\hat{w}'_i>\hat{w}_i$, $\hat{a}'_i \le \hat{a}_i$, and $\hat{d}'_i \ge \hat{d}_i$, assuming other bids are fixed. This condition can be easily verified. So, Auction~1 is DSIC.
\end{proof}

\begin{algorithm}[t]
    \caption{Critical Price Calculation for Requests}\label{alg:cri}
    In each time slot $t$:
    \begin{algorithmic}[1]
    \FORALL{$i \in D$}
		\IF{$i$ is scheduled by Algorithm~\ref{alg:on}}
			
			\STATE $w_{low}\leftarrow 0$
			\STATE $w_{high}\leftarrow w_i$
			\WHILE{$w_{low}<w_{high}$}

				\STATE $cp_i\leftarrow \frac{w_{low}+w_{high}}{2}$	
				
				\STATE Run Algorithm~\ref{alg:on} with the valuation of request $i$ updated by $cp_i$ from $t={a}_i$ to ${d}_i$
				\IF{$i$ is scheduled}
					\STATE $w_{high}\leftarrow cp_i$
				\ELSE
					\STATE $w_{low}\leftarrow cp_i$
				\ENDIF
				
			\ENDWHILE
			\STATE Output $cp_i$ as the critical price for $i$
			
		\ENDIF
    \ENDFOR

    \end{algorithmic}
    \vspace{-0.2em}
\end{algorithm}

By the definition of critical price, we propose Algorithm~\ref{alg:cri} that applies binary search to find the critical price for requests scheduled by Algorithm~\ref{alg:on}. Algorithm~\ref{alg:cri} runs in each slot $t$ when there are requests scheduled. In the binary search from Lines~6-12, scheduling decisions must be remade from $t=a_i$ to $d_i$ with $w_i$ updated by the new value of $w_i$ (Line~7) till the critical price is found.

\subsection{Non-negative Revenue for the Operator}\label{sec:revenue}

Our objective in Problem~(A) is to maximize the social welfare without considering the revenue at the operator side. However, for an actual business model to be viable, it is important that the revenue of the operator is taken into account. The revenue is composed of two parts: Payments collected from the SUs by serving their requests and the penalty paid for causing collisions. Using $Q(1-P_0(k))/P_0(k)$ as $\theta(k)$ in Algorithm~\ref{alg:on}, the operator may get a negative revenue, which means the sum of payments by SUs does not exceed the penalty paid to PUs out of its control. To overcome this problem, we introduce a \emph{reservation price} $q$, which is a constant for fixed channel related parameters. We further reset the value of $\theta(k)$ to be $q$ in Algorithm~\ref{alg:on} and apply Auction~1.




We interpret the reservation price as the expected cost associated with each request served by channels in $T_2$.
Hence, the revenue of the operator will always be non-negative if at least the reservation price is paid by the SU for getting its request served. Next, we show the form of the reservation price for the special case in Proposition~\ref{prop:cri}.

\begin{proposition}
\label{prop:cri}
If $|T_1|=0$ and channels in $T_2$ are homogeneous, then a reservation price $q_0=Q(1-P_0)/P_0$ leads to non-negative revenue for the operator.
\end{proposition}
\begin{proof}
The revenue of the operator is expected payment collected from SUs by serving their requests minus the cost. The payment of SUs is at least the reservation price $Q(1-P_0)/P_0$. By Lemma~\ref{lem:cost}, we know that the expected cost is $Q(1-P_0)/P_0$ as well. Hence, the expected revenue is non-negative.
\end{proof}

\ignore{Note that, $q_0$ has been chosen as the value of $\theta(k)$ in the greedy online algorithm in Section~\ref{sec:on_wo} for maximizing social welfare only.}
When there are heterogeneous channels, we try to find the average expected cost of serving a request and set it as the reservation price to guarantee a non-negative revenue. \ignore{Note that $T_1$ channels have no cost and $T_2$ channels have different positive costs for serving requests.} Recall that $c_k$ is the expected cost of serving a request on channel $k$ by Lemma~\ref{lem:cost}. Assume that channels have been sorted by a non-decreasing order of $c_k$. For ease of illustration, we index channels in $T_1$ from $1$ to $|T_1|$, and channels in $T_2$ from $|T_1|+1$ to $|T_1|+|T_2|$. Let $n'_j$ denote the expected fraction of requests served by channel $j$ for a given set of requests. Then, the average expected cost per request is $q'\overset{\triangle}=\sum\limits_{j\in T_1\cup T_2} {c_j n'_j } $. We would like to have an estimate of $q'$ that is independent of the request set. To this end, let $v_k$ denote the probability that the channel $k$ is viewed as idle and it is really idle. That is, $v_k=\pi_1(k)$ if $k\in T_1$, and $v_k= P_I(k)P_0(k)$ if $k\in T_2$. Let $m_k\overset{\triangle} = \frac{v_k}{\sum\limits_{l\in T_1\cup T_2}{v_l}}$. We use $m_k$ as the estimated fraction of requests served by channel $k$. Then we let $q_1 \overset{\triangle}= \sum\limits_{j\in T_1\cup T_2} {c_j m_j }$, which we use to estimate the average expected cost per request and set it as the reservation price. We would like to show that $q_1 \geq q'$ and hence using $q_1$ as a reservation price, a non-negative expected revenue is obtained. We start with Lemma~\ref{lem:suf_res} that provides a sufficient condition for $q_1 \geq q'$.

\begin{lemma}
If $\frac{m_j}{m_{j+1}} \le \frac{n'_j}{n'_{j+1}}$ for all $j$, then $q_1 \ge q'$.
\label{lem:suf_res}
\end{lemma}
\begin{proof}
We calculate $q_1-q'=\sum\limits_{j\in T_1\cup T_2}c_j (m_j-n'_j)$. In the following, we will show that there exists $i$ such that for all $j\le i$ we have $m_j\le n'_j$ and for all $k> i$ we have $m_k> n'_k$. Since $\frac{m_j}{m_{j+1}} \le \frac{n'_j}{n'_{j+1}}$ for all $j$, it is easy to see that: if $m_j\ge n'_j$, then $m_k\ge n'_k$ by multiplying $\frac{m_{j+1}}{m_j}\cdots \frac{m_{k}}{m_{k-1}}$ and $\frac{n'_{j+1}}{n'_j}\cdots \frac{n'_{k}}{n'_{k-1}}$, respectively, on both sides. Then we can find such $i$. We divide $q_1-q'$ into two parts:
\begin{equation}
q_1-q'=\sum\limits_{j\le i}c_j (m_j-n'_j)+\sum\limits_{k> i}c_k (m_k-n'_k)
\label{eq:sum}
\end{equation}

If $i=|T_1|+|T_2|$, $q_1-q' = \sum\limits_{j \in T_1\cup T_2 }c_j (m_j-n'_j) \geq (\max\limits_{j \in T_1\cup T_2}{c_j})\Big(\sum\limits_{j\in T_1\cup T_2}{m_j}-\sum\limits_{j\in T_1\cup T_2}{n'_j}\Big)=(\max\limits_{j \in T_1\cup T_2}{c_j}) (1 - 1) = 0$. If $i=0$, then all terms in $q_1-q'$ are positive. Next we consider the case where neither sums in Equation~(\ref{eq:sum}) has no terms. Since all terms in the first term in the sum are non-positive and all terms in the second term in the sum are positive, we obtain
\[q_1-q'\ge(\max\limits_{j\le i}{c_j})\sum\limits_{j\le i}{(m_j-n'_j)}+(\min\limits_{k> i}{c_k})\sum\limits_{k> i}{(m_k-n'_k)}\]
\[\overset{(a)}\ge (\max\limits_{j\le i}{c_j})\Big(\sum\limits_{j\le i}{(m_j-n'_j)}+\sum\limits_{k> i}{(m_k-n'_k)}\Big)\]
\[=(\max\limits_{j\le i}{c_j})\Big(\sum\limits_{j\in T_1\cup T_2}{m_j}-\sum\limits_{j\in T_1\cup T_2}{n'_j}\Big)=0,\]

\noindent where (a) is by the assumption that $c_1\le \cdots \le c_{|T_1|+|T_2|}$. Hence $q_1\ge q'$ holds.
\end{proof}

Based on Lemma~\ref{lem:suf_res}, we claim that a reservation price of $q_1$ results in a non-negative revenue for the operator.

\begin{proposition}
The operator achieves a non-negative expected revenue with reservation price $q_1$ when $H$ is large enough.
\label{prop:gen_res}
\end{proposition}
\begin{proof}
It suffices to show that $q_1 \ge q'$. Consider any sample path. Without loss of generality, consider the first two channels in the sorted list. Let $n_1$ and $n_2$ denote the number of requests served by channels $1$ and $2$, respectively. Let $s_i$ denote the number of time slots that are sensed in the interval between $(i-1)$th and $i$-th requests served by channel $1$, and define $b_i$ similarly for channel $2$. Let $A$ denote the total number of time slots between $0$ and $H$ that are not sensed for channel $1$ (because there is not enough requests or there is no benefit to sense it), and $A'$ the number of time slots after the last request is served by channel $1$. Define $B$ and $B'$ similarly for channel $2$. Note that by the ordering of channels, when there is only one request in the system, and both channels are available, channel 1 will be used. It follows that $A \le B$. We then have $H = \sum {s_i} + A + A' = \sum {b_i} + B + B'$. Therefore, $H = E(\sum{s_i} + A + A') = E(n_1)/m_1 + E(A) + E(A')$ (by geometric distribution) and $H= E(n_2)/m_2 + E(B) + E(B')$.  Note that $E(A')/H\to 0$ and $E(B')/H \to 0$ when $H \to \infty$. Therefore $\frac{n'_1}{n'_2} =  \frac{E(n_1)}{E(n_2)}=\frac{[H - E(A)]m_1}{[H - E(B)]m_2} \ge \frac{m_1}{m_2}$ since $H-E(A) \geq H-E(B)$. It then follows that $q_1 \geq q'$ by Lemma~\ref{lem:suf_res}. Hence, a reservation price of $q_1$ leads to a non-negative revenue at the operator.
\end{proof}

\emph{Remark:} With a lower reservation price, the online algorithm tends to access the channels more aggressively, thus, the revenue of operator is harmed. On the other hand, a higher reservation price prevents more requests from accessing the channels, which harms the social welfare and further affects the revenue, as well. We will evaluate the tradeoff between social welfare and revenue by setting different reservation prices in Section~\ref{sec:simu}.

\section{Numerical Result}
\label{sec:simu}

In this section, we evaluate the performance of the greedy online algorithm (Algorithm~\ref{alg:on}) and the tradeoff between social welfare and revenue for different reservation prices. We first show the competitive ratio of the greedy online algorithm with different channel settings and request related parameters, respectively. We then apply Auction~1 with varying reservation prices and show the performances of social welfare and revenue.

\subsection{Simulation Setting}
We let the arrivals $a_i$ of requests follow a Poisson distribution and the duration $d_i-a_i$ of the requests follow an exponential distribution. The valuations follow a uniform distribution in [1,15]. We choose $Q=10$, the penalty per collision, comparable to the valuations in all our simulations. We fix the number of requests as $20$, and the inter-arrival mean as $3$ slots, and vary the mean of request duration to adjust the density of requests. Given the means of inter-arrival and request durations, we generate $50$ groups of requests and compare the average for the metrics we consider. We generate the channel availabilities in each time slot based on our assumption that channel states follow i.i.d Bernoulli distribution and $100$ samples of channel realizations are taken for our simulations. The channel parameters we use will be introduced in Section~\ref{subsec:perf_on}.

\subsection{Performance of Greedy Online Algorithm}
\label{subsec:perf_on}
\begin{figure}[!t]
\centering
\subfigure[\small Homogeneous $T_2$ channels.]{
\includegraphics[scale=0.24]{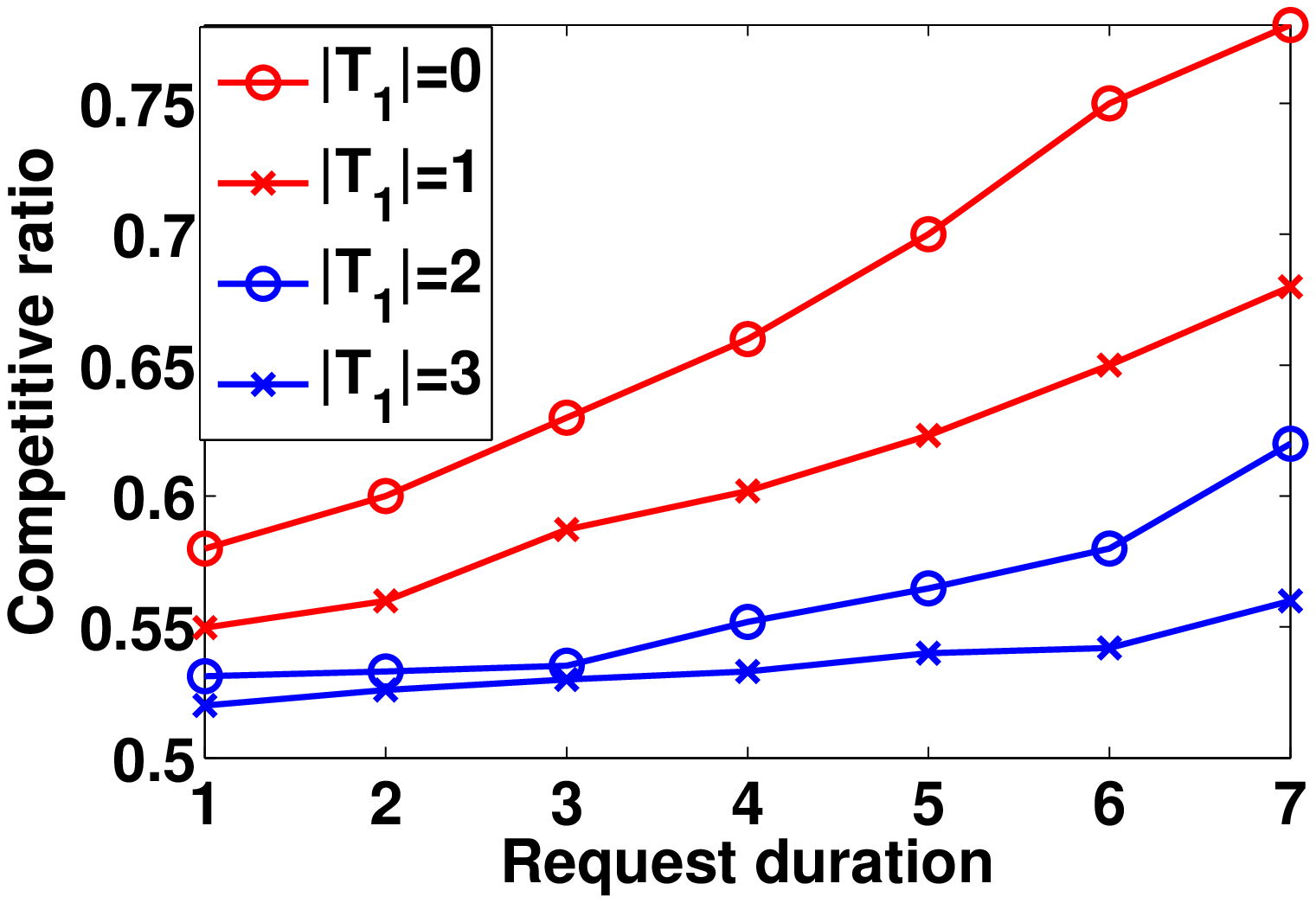}
\label{fig:homo_dur}
}
\hfil
\subfigure[\small Heterogeneous $T_2$ channels.]{
\includegraphics[scale=0.25]{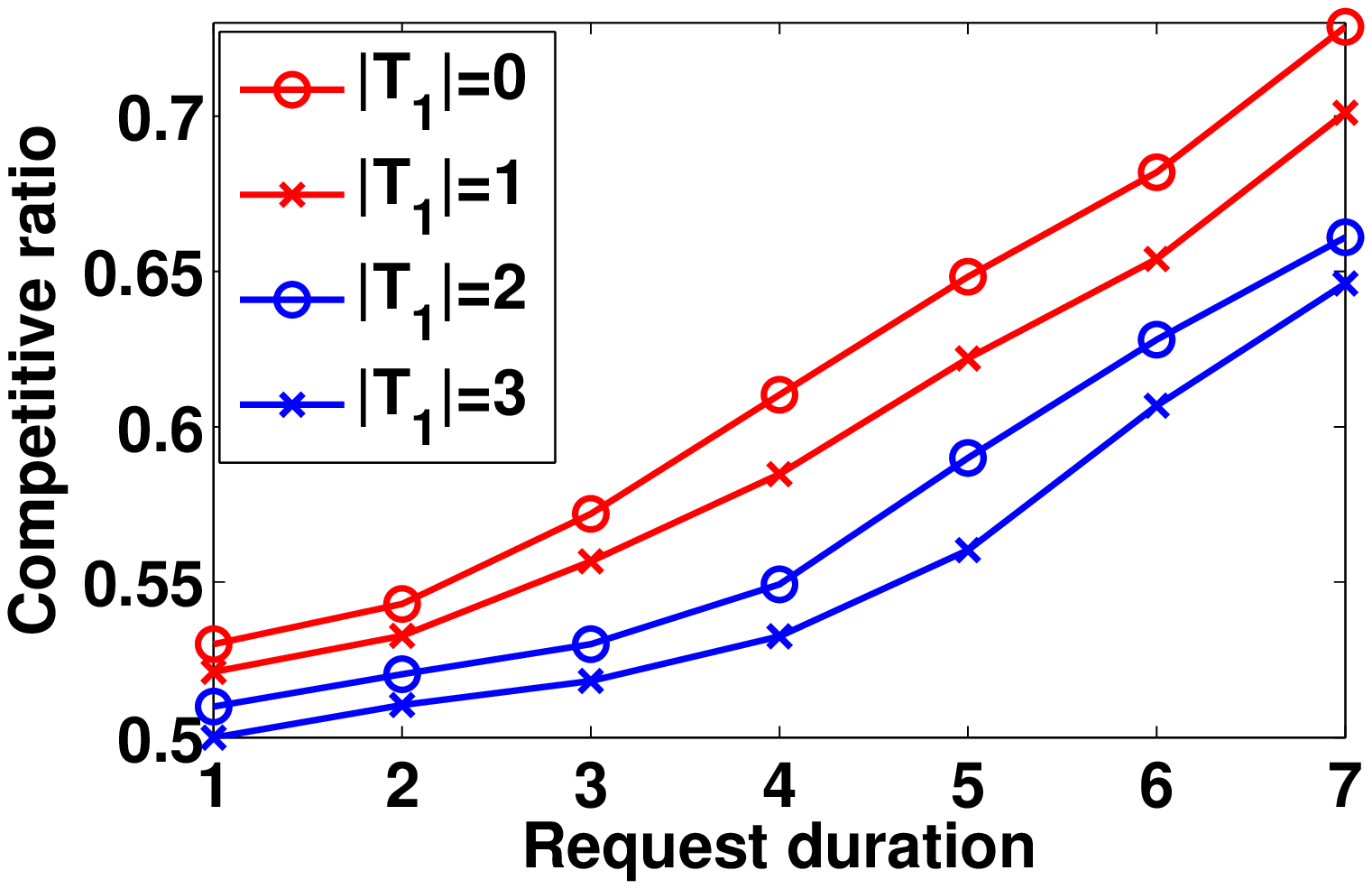}
\label{fig:heter_dur}
}

\vspace{-0.5em}
\caption{Performance of online algorithm versus offline algorithm over various request duration means with homogeneous and heterogeneous $T_2$ channels ($|T_2| = 3$), respectively.}
\vspace{-1.5em}
\label{fig:ratio}
\end{figure}

In Figure~\ref{fig:homo_dur}, we compare the performance of Algorithm~\ref{alg:on} with that of Algorithm~\ref{alg:off} when there are three homogeneous $T_2$ channels in the system with ${\pi}_2=0.6324$, $P_m=0.2218$, $P_f=0.6595$ and various number of $T_1$ channels. The $y$-axis denotes the achieved competitive ratio, i.e., the ratio between the social welfare of the online algorithm and that of the optimal offline algorithm. When $|T_1|=1$, we set ${\pi}_1=0.5058$; When $|T_1|=2$, we set ${\pi}_1(1)=0.8147$ and ${\pi}_1(2)=0.1270$. We observe that the performance of Algorithm~\ref{alg:on} degrades as $|T_1|$ increases, independent of the request duration mean. With a high number of $T_1$ channels, a wrong decision made by the greedy online algorithm to schedule a request affects the performance more. Also, the greedy online algorithm serves requests of a larger density better than requests of a smaller density. When the system is overloaded with requests, even the optimal offline algorithm can not satisfy all requests. Thus, those with larger valuations tend to be chosen, as in the greedy online algorithm. All ratios plotted are strictly above $\frac{1}{2}$, even for those with $|T_1|\neq 0$.

In Figure~\ref{fig:heter_dur}, we evaluate the performance of Algorithm~\ref{alg:on} with heterogeneous $T_2$ channels. We use the same $T_1$ channel parameters as in the homogeneous case. The parameters related to $T_2$ channels are as follows: ${\pi}_2(1)=0.9134$, ${\pi}_2(2)=0.6324$, ${\pi}_2(3)=0.0975$, $P_m(1)=0.1419$, $P_f(1)=0.7922$, $P_m(2)=0.2218$, $P_f(2)=0.6595$, $P_m(3)=0.6557$, $P_f(3)=0.2157$. We observe similar results as in Figure~\ref{fig:homo_dur}: Algorithm~\ref{alg:on} performs better with fewer $T_1$ channels and denser requests. Again all ratios are above $\frac{1}{2}$.

\subsection{Tradeoff between Social Welfare and Revenue}
\begin{figure}[!t]
\centering
\subfigure[\small Homogeneous $T_2$ channels with $|T_1|=0$.]{
\includegraphics[scale=0.25]{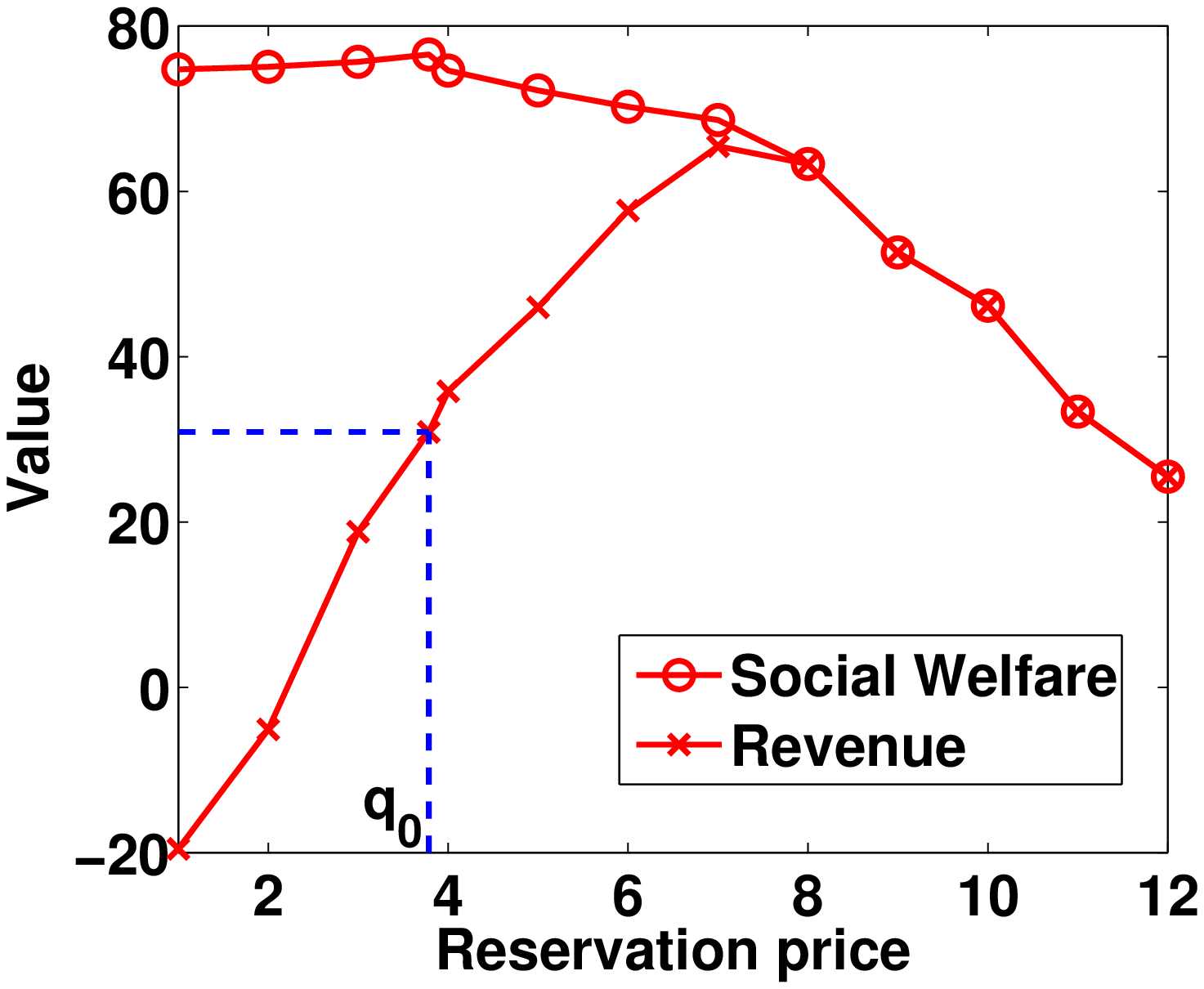}
\label{fig:homo_tradeoff}
}
\hfil
\subfigure[\small Heterogeneous $T_2$ channels with $|T_1|=1$.]{
\includegraphics[scale=0.25]{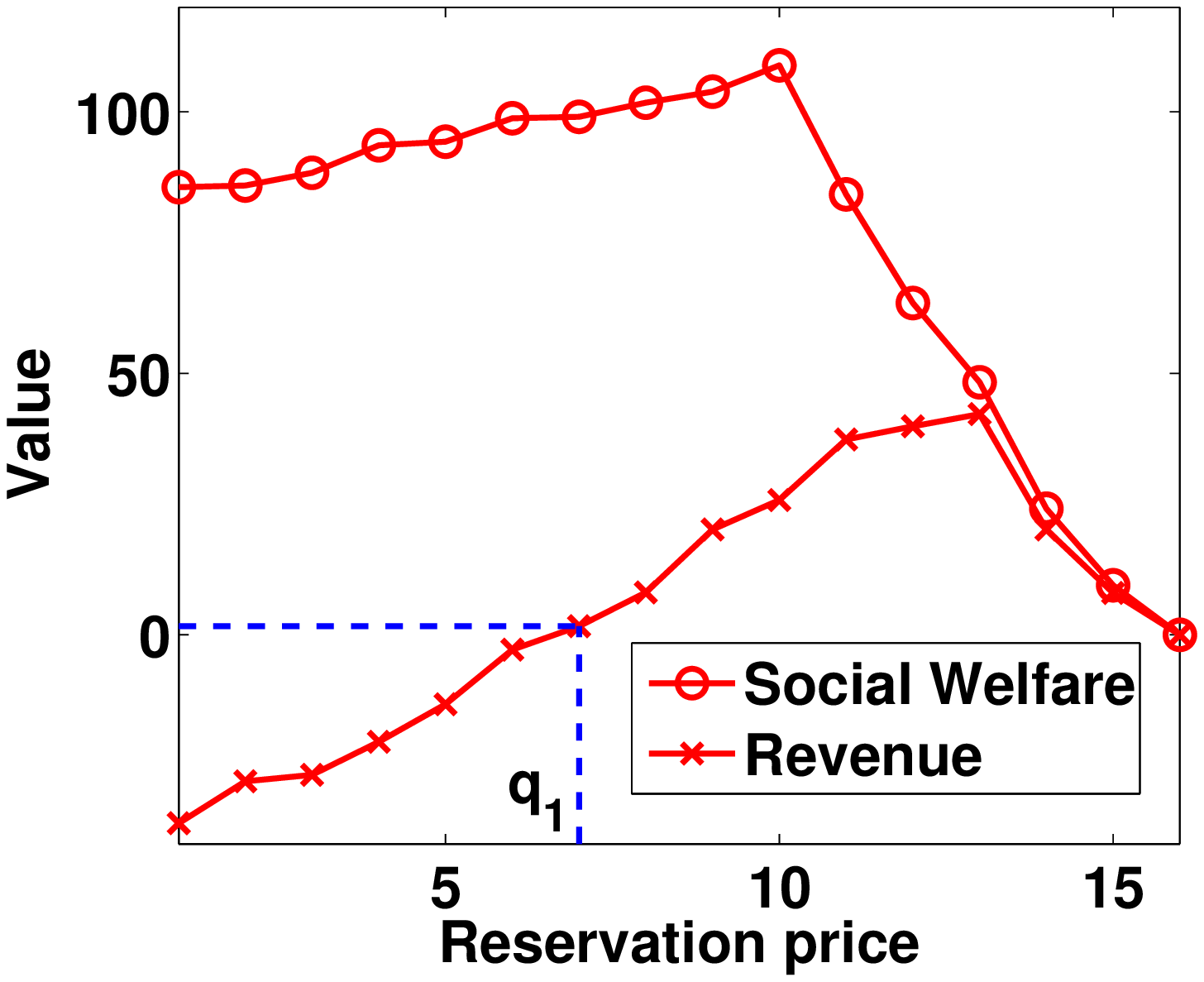}
\label{fig:heter_tradeoff}
}

\vspace{-0.5em}
\caption{Tradeoff between social welfare and revenue over reservation price with homogeneous and heterogeneous $T_2$ channels ($|T_2|=3$), respectively.}
\vspace{-1.5em}
\label{fig:tradeoff}
\end{figure}

We now study the tradeoff between social welfare and revenue generated by Auction~1. In Figure~\ref{fig:tradeoff}, we vary the values of reservation price. Note that it is now a constant over requests given the channel related parameters, which are the same as in Section~\ref{subsec:perf_on}. We first show the tradeoff in a system with homogeneous $T_2$ channels and no $T_1$ channels in Figure~\ref{fig:homo_tradeoff}. Both social welfare and revenue first increase and then decrease as the reservation price increases. At a low reservation price ($<3$), the payment collected cannot recover the expected cost and hence the average revenue becomes negative. A low reservation price may also hurt social welfare by our necessary condition for serving requests (Proposition~\ref{prop:nec}). On the other hand, when the reservation price is too high, fewer requests will be accepted, which hurts both social welfare and revenue. \ignore{though the change point of the former is smaller than that of the latter. This matches our intuition that social welfare will be affected in a faster way when the reservation price increases, and revenue will be harmed when the reservation price is high, which prevents requests from being served. Thus, less payment will be collected.} We note that when the reservation price is $q_0 = Q(1-P_0)/P_0=3.8$, a non-negative revenue is obtained, which is consistent with Proposition~\ref{prop:cri}.\ignore{A reservation price as low as $3$ also results in a non-negative revenue in the figure.} At a very high reservation price ($\ge 8$), the social welfare and the revenue converge, where the payment actually becomes the same as the valuation for requests served. Note that the revenue never exceeds the social welfare by the definition of critical price.

In Figure~\ref{fig:heter_tradeoff}, we show the tradeoff in a system with $T_1$ and heterogeneous $T_2$ channels. The trend of social welfare and revenue is similar to that in Figure~\ref{fig:homo_tradeoff}. \ignore{The main difference lie in that the change point of social welfare is $10$, larger than $4$ in Figure~\ref{fig:homo_tradeoff}, and the change point of revenue is $13$, larger than $7$ in Figure~\ref{fig:homo_tradeoff}.The revenue drops to $0$ when the reservation price is $15$ since the maximum valuation we set is $15$. Thus, no requests will be served in the auction after that.} Note that with the reservation price $q_1 = 6.9$ (defined in Section~\ref{sec:inc}), the revenue obtained is right above 0, which is consistent with Proposition~\ref{prop:gen_res} and also shows that $q_1$ is nearly a tight upper bound of the expected cost for this case. \ignore{Comparing these two figures, we find that the ratio of the maximum revenue versus the maximum social welfare is much lower in the heterogeneous case, which may imply that a fixed reservation price for all channels is less efficient in the heterogeneous case.} \ignore{On the hand, the revenue heterogeneity in channels (including both $T_1$ and $T_2$ since they are different types) contributes to social welfare. However, revenue is harmed by the heterogeneity as reflected in the observation that the highest revenue in Figure~\ref{fig:heter_tradeoff} is only $42.2$ while it is $65.4$ in Figure~\ref{fig:homo_tradeoff}.} 
\section{Conclusion}
In this paper, we study the joint sensing and spectrum allocation problem for serving secondary users in cognitive radio networks with the objective of maximizing the social welfare. Our problem formulation takes into account both spectrum uncertainty and sensing inaccuracy, which enables dynamic spectrum access at small time scales. Using only channel statistics and real time channel states, we develop an optimal solution for serving a given set of spectrum requests with various time elasticity. We further propose an online algorithm, which does not require future information on the arrival process, and achieves a comparable performance as the offline algorithm. In addition, we show that the online algorithm together with a payment scheme achieves incentive compatibility for the SUs and a non-negative revenue for the operator. There are several open problems to be solved. First, in practice, a more flexible form of spectrum requests will be desirable. For instance, a request may ask for multiple chunks that may or may not be preemptive. Extending the current offline and online algorithms to this more general setting will be part of our future work. Second, we plan to extend the problem formulation by including the notion of spatial spectrum reuse in addition to the time dimension considered in the paper. Third, we plan to relax the assumption on the i.i.d Bernoulli channels by considering correlated channels, which involves solving an {\it exploration vs. exploitation} problem in the context of an auction.
\label{sec:con}

\bibliographystyle{abbrv}
\bibliography{main}

\begin{thebibliography}{10}

\bibitem{competitive-analysis}
A.~Borodin and R.~El-Yaniv.
\newblock {\em Online Computation and Competitive Analysis}.
\newblock Cambridge University Press, 1998.

\bibitem{Chen}
L.~Chen, S.~Iellamo, M.~Coupechoux, and P.~Godlewski.
\newblock An auction framework for spectrum allocation with interference
  constraint in cognitive radio networks.
\newblock In {\em Proc. of IEEE Infocom}, pages 794--802, Piscataway, NJ, USA,
  2010. IEEE Press.

\bibitem{Deek}
L.~B. Deek, X.~Zhou, K.~C. Almeroth, and H.~Zheng.
\newblock To preempt or not: Tackling bid and time-based cheating in online
  spectrum auctions.
\newblock In {\em Proc. of IEEE Infocom}, pages 2219--2227, 2011.

\bibitem{Wang}
M.~Dong, G.~Sun, X.~Wang, and Q.~Zhang.
\newblock Combinatorial auction with time-frequency flexibility in cognitive
  radio networks.
\newblock In {\em Proc. of IEEE Infocom}, pages 2282 --2290, march 2012.

\bibitem{Gao}
L.~Gao, Y.~Xu, and X.~Wang.
\newblock Map: Multiauctioneer progressive auction for dynamic spectrum access.
\newblock {\em IEEE Transactions on Mobile Computing}, 10:1144--1161, 2011.

\bibitem{Gop}
A.~Gopinathan, Z.~Li, and C.~Wu.
\newblock Strategyproof auctions for balancing social welfare and fairness in
  secondary spectrum markets.
\newblock In {\em Proc. of IEEE Infocom}, pages 3020--3028. IEEE, 2011.

\bibitem{haj}
M.~T. Hajiaghayi, R.~D. Kleinberg, M.~Mahdian, and D.~C. Parkes.
\newblock {Online auctions with re-usable goods}.
\newblock In {\em Proceedings of the 6th ACM conference on Electronic
  commerce}, EC '05, pages 165--174, New York, NY, USA, 2005. ACM.

\bibitem{sensing-survey}
{I. F. Akyildiz, B. F. Lo, and R. Balakrishnan}.
\newblock Cooperative spectrum sensing in cognitive radio networks: A survey.
\newblock {\em Physical Communication}, 4:40--62, mar 2011.

\bibitem{li-single-channel}
S.~Li, Z.~Zheng, E.~Ekici, and N.~Shroff.
\newblock Maximizing system throughput by cooperative sensing in cognitive
  radio networks.
\newblock In {\em Proc. of IEEE Infocom}, mar 2012.

\bibitem{li-multi-channel}
S.~Li, Z.~Zheng, E.~Ekici, and N.~Shroff.
\newblock Maximizing system throughput using cooperative sensing in
  multi-channel cognitive radio networks.
\newblock In {\em Proc. of IEEE CDC}, dec 2012.

\bibitem{Nisan}
N.~Nisan, T.~Roughgarden, E.~Tardos, and V.~V. Vazirani.
\newblock {\em Algorithmic Game Theory}.
\newblock Cambridge University Press, New York, NY, USA, 2007.

\bibitem{Puterman}
M.~L. Puterman.
\newblock {\em Markov Decision Processes: Discrete Stochastic Dynamic
  Programming}.
\newblock John Wiley \& Sons, Inc., New York, NY, USA, 1st edition, 1994.

\bibitem{Xu}
P.~Xu and X.-Y. Li.
\newblock Online market driven spectrum scheduling and auction.
\newblock In {\em Proc. of ACM CoRoNet}, pages 49--54, New York, NY, USA, 2009.
  ACM.

\bibitem{Zhou}
X.~Zhou, S.~G, S.~Suri, and H.~Zheng.
\newblock ebay in the sky: Strategy-proof wireless spectrum auctions.
\newblock In {\em Proc. of ACM MobiCom}, 2008.

\end{thebibliography}

\end{document}